# 6 Evidence against non-cosmological redshifts of QSOs in SDSS data

**Sumin Tang**

Harvard-Smithsonian Center for Astrophysics, 60 Garden Street, Cambridge, MA 02138, USA

**Shuang Nan Zhang**

Department of Physics and Center for Astrophysics, Tsinghua University, Beijing 100084, China

## Abstract

In the unusual intrinsic QSO redshift models, QSOs are ejected by active galaxies with periodic non-cosmological reshifts, thus QSOs are generally associated with active galaxies, and certain structures will be revealed in the QSO redshift distribution. As the largest homogeneous sample of QSOs and galaxies, SDSS data provide the best opportunity to examine this issue. We review the debates on this issue, focused on those based on SDSS and 2dF data, and conclude that there is no strong connection between foreground active galaxies and high-redshift QSOs. The existence of two dips in the SDSS QSO redshift distribution at $z=2.7$ and $3.5$ has recently re-ignited those controversial debates on the origin of QSO redshift. It also turned out that both dips are totally caused by selection effects and after selection effects have been corrected, the two dips disappear and no structure in the redshift distribution of SDSS DR5 sample. These results support that the reshifts of QSOs are cosmological.

E-mails: stang@cfa.harvard.edu (ST); zhangsn@tsinghua.edu.cn (SNZ)

# Introduction

Whether quasi-stellar objects (QSOs) are non-cosmological objects is an old debate, dating back to late 1960s [1-3], shortly after the first discovery of QSO in 1963 [4]. Some evidence has been claimed to suggest the intrinsic non-cosmological redshift hypothesis, in which QSOs are ejected by nearby peculiar galaxies with much larger redshifts than their parent galaxies, and QSO redshifts have nothing to do with their distances [1-3, 5-7].

Besides the association of QSOs with nearby galaxies, the abnormal distribution of QSO redshifts with preferred redshifts is viewed as another supporting evidence of non-cosmological redshift models, because if strongly preferred redshifts do exist, they cannot be explained in the standard cosmological redshift model. Such non-cosmological redshift is called intrinsic redshift and is produced by unknown mechanisms [2, 3, 8-14].

The above non-cosmological redshift hypothesis, if true, the consequences will be enormous and it will overthrow our current understanding of cosmology and the nature of QSOs. Most previous studies supporting non-cosmological redshifts of QSOs used rather small heterogeneous samples (except [10,12,34] which we will discuss more in Section 2), which suffered from biases as well as selection effects [33, 15]. Statistical arguments either supporting non-cosmological redshift models or against non-cosmological redshift models were not strong enough before the advents of 2dF and SDSS [16,17].

As the largest homogeneous samples of QSOs and galaxies, SDSS and 2dF data provide the best opportunity to examine this issue. Here we review the non-cosmological redshift models in the literature, and summarize the evidence against non-cosmological redshift models in SDSS and 2dF data [18-20].

# 1. Non-Cosmological Redshift Models of QSOs

Shortly after the first discovery of QSO 3C 273 in 1963 [4], Arp in 1966 found some radio sources, including 5 known QSOs, are close to peculiar galaxies with separation of several degrees on the sky, and based on this phenomenon, he proposed that QSOs are ejected by nearby peculiar galaxies thus not at cosmological distances [1]. However, no reasonable mechanism could self-consistently explain the origin of the large 'non-cosmological' redshifts of QSOs [1,13,14]. In 1967, Burbidges found the absorption line redshifts are close to $z=1.95$ in a group of 7 QSOs [2]. They proposed that 1.95 is an intrinsic non-cosmological redshift of QSOs, while the origin remains unknown. In 1968, Cowan found a periodicity of 0.167 in QSO redshifts based on a sample of 116 QSOs [3].

One main original drive of the non-cosmological redshift models is that it could solve the energy problem of QSOs, i.e. the enormous energy output (if the redshift is cosmological) from a very small region, which was a big challenge at that time, where mechanisms other than nuclear burning in stars are needed [2,13,21]. In 1969, Lynden-Bell pointed out that energy release from gravitational infall is more efficient than nuclear burning, hence the high QSO luminosity problem could be solved by accretion onto the supermassive black hole in the center of a galaxy [22,23].

Later on, several groups of people continued studying on the non-cosmological redshifts of QSOs and reported anomalies to suggest this non-cosmological QSOhypothesis [5-12]. The anomalies could be divided into two groups. The first one is the visual (line of sight) association of QSOs with nearby galaxies [e.g. 24,25], and the second one is structures, including periodicities, peaks or dips, in the redshift distribution of QSOs [8,9,14]. The current status of non-cosmological redshift study is well reviewed by [26].

For the anomalous redshift distributions, there are two models discussed in the literature that predict exact values for the preferred redshifts [19]. One of these is the Karlsson formula, which suggests a periodicity of 0.089 in $\log(1+z_{eff})$ with peaks lying at 0.061, 0.30, 0.60, 0.96, 1.14, 1.96, and so on [5,7-11], where the effective redshift $z_{eff}$ is the redshift of the QSO measured relative to the nearby galaxy, which is defined as

$$1 + z_{eff} = (1+ z_Q)/(1+ z_G) \tag{1}$$

where $z_Q$ is the observed quasar redshift and $z_G$ is the redshift of the associated galaxy that is assumed to be the ejecting galaxy. In their model, quasars are ejected by active galaxies and their putative parent galaxies are generally much brighter than their quasar offsprings [10], and the typical projected association separation is about 200 kpc [7,11].

The other model, namely the decreasing intrinsic redshift model (DIR model), was proposed by Bell [12], where the QSO intrinsic redshift equation is given by

$$z_{iQ} = z_f (N - M_N) \tag{2}$$

where $z_f = 0.62$ is the intrinsic redshift constant, $N$ is an integer, and $M_N$ is a function of $N$ and another quantum number $n$. In the DIR model, galaxies are produced continuously through the entire age of the universe, and QSOs are ejected from the nuclei of active galaxies at the early stage of the evolution of galaxies.

## 2. Evidence against Non-Cosmological Redshifts of QSOs in SDSS and 2dF data

As the largest homogeneous samples of QSOs and galaxies, SDSS and 2dF data provide the best opportunity to examine this issue, which could overcome the biases and selection effects from small and heterogeneous samples in most previous studies. Here we summarize the debates on non-cosmological QSO redshift models based on SDSS and 2dF data, and we show that observational evidence is against non-cosmological redshifts.

### 2.1 Debates on QSO associations with nearby active galaxies

Lots of anomalies in the associations of QSOs with nearby active galaxies have been reported in the literature, as discussed in Section 1. However, most of them are based on a very small sample or selected only a few galaxies from a large sample, therefore severely suffered from chance alignments. López-Corredoira and Cutiérrez searched for anisotropy in the SDSS QSO distribution around nearby edge-on spiral galaxies and reported a tentative 3.5-sigma detection, which is very interesting, but needs further confirmation as well as removals of extinction and gravitational lensing effects [34].

In 2005, we tested the QSO associations with nearby active galaxies using SDSS data [19]. We used the SDSS Data Release 1 (DR1) QSO catalog [27,28] and the New York University Value-Added Galaxy Catalog (NYU-VAGC) [29]. For reliability in the derived redshifts, we considered only those QSOs in the range of z > 0.4 and galaxies in the range of 0.01< z < 0.2 with the highest plate quality labeled as ''good'' and with no redshift warning. This quality control leaves a total of 190,591 galaxies and 15,747 QSOs in the sample. We compared the data with the ejection models, especially Bell's DIR model. We found that the sample is fully consistent with randomly distributed QSOs and galaxies, and incompatible with ejection models, as shown in Figures 1 and 2. Results of $\chi^2$ tests are given in Table 1, which shows that at 3-sigma confidence level for both distributions of projected separation distance and redshift distribution of active galaxies in pairs with QSOs, the data are consistent with random distributions, but inconsistent with the ejection hypothesis. More details and discussions are in [19].

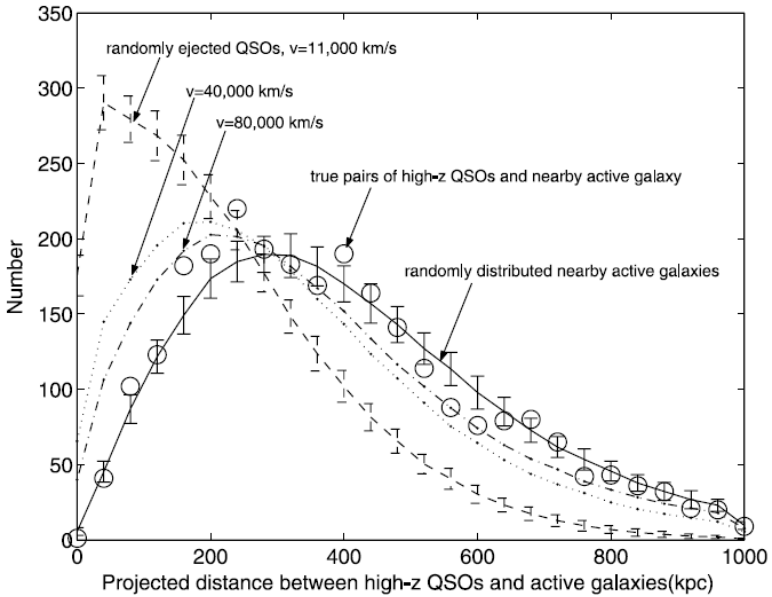

**Figure 1.** Distribution of projected distance between 2604 high-z QSOs (2.4 < z < 4.8) and their paired nearby active galaxies in NYU-VAGC. The circles represent ''true'' pairs, i.e., pairs found in the data, but not necessarily physical pairs. The solid line with error bars is the average of 200 simulations of QSOs and randomly distributed galaxies. Averages of 200 simulations of randomly ejected QSOs and active galaxies are also presented, in which QSOs are produced by ejection from randomly chosen galaxies with a uniformly distributed age in $0$–$10^8$ yr and three different velocities: dashed line with error bars, 11,000 km/s; dotted line with points, 40,000 km/s; and dash-dotted line with points, 80,000 km/s. (This figure is taken from [19].)

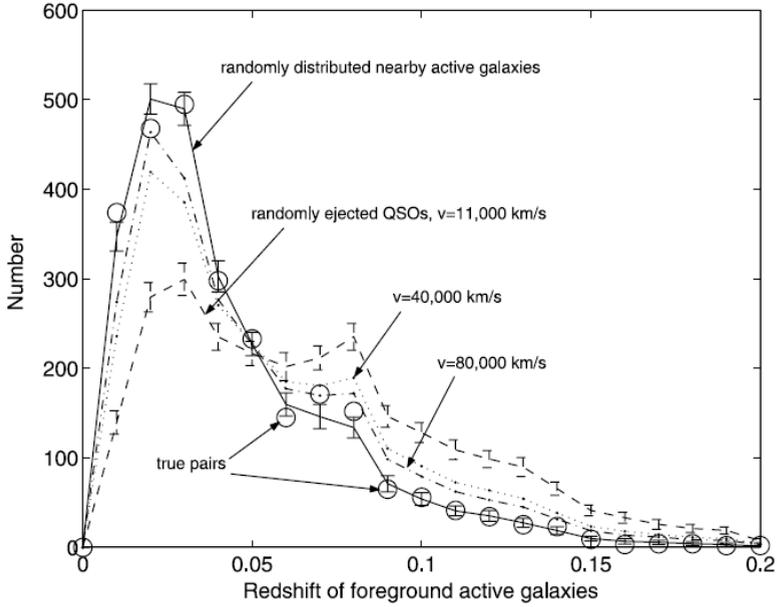

**Figure 2.** Distribution of 2604 foreground active galaxies that have at least one high-z QSO within a projected distance of 1 Mpc in NYU-VAGC. The circles represent ''true'' pairs. Others are as in Fig. 7. (This figure is taken from [19].)

TABLE 1

RESULTS OF $\chi^2$ TESTS FOR THE DISTRIBUTION OF PROJECTED SEPARATION DISTANCE AND THE REDSHIFT DISTRIBUTION OF ACTIVE GALAXIES IN PAIRS WITH HIGH-z QSOs BETWEEN "TRUE" PAIRS AND SIMULATIONS

| Simulations (1) | $\chi^{2a}$ (2) | $\chi^2/N^a$ (3) | $p^a$ (4) | $\chi^{2b}$ (5) | $\chi^2/N^b$ (6) | $p^b$ (7) |
|---|---|---|---|---|---|---|
| Randomly distributed galaxies | 38.456 | 1.6720 | 0.023 | 18.818 | 0.9904 | 0.47 |
| Ejected QSOs: | | | | | | |
| $v = 11{,}000$ km s$^{-1}$ | 2560.3 | 111.3174 | $<10^{-10}$ | 994.10 | 52.3211 | $<10^{-10}$ |
| $v = 40{,}000$ km s$^{-1}$ | 227.12 | 9.8748 | $<10^{-10}$ | 265.31 | 13.9637 | $<10^{-10}$ |
| $v = 80{,}000$ km s$^{-1}$ | 85.496 | 3.7172 | $4.0 \times 10^{-9}$ | 144.37 | 7.5984 | $<10^{-10}$ |

[a] Cols. (2) and (3) are for the distribution of projected separation distance. Pairs with projected separation distance less than 60 kpc (first and second points in Fig. 7) are not used in the tests to avoid the SDSS 55″ fiber constraint, and all simulations are normalized to get a minimum $\chi^2$. The degrees of freedom are $N = 23$.
[b] Cols. (4)–(7) are for the redshift distribution of active galaxies in pairs with high-z QSOs. All of the first zero points in Fig. 8 are not used in the tests, and all simulations are normalized to get a minimum $\chi^2$. The degrees of freedom are $N = 19$.

**Table 1.** This table is taken from [19].

## 2.2 Debates on Periodicities, Peaks or Dips in QSO Redshifts

In 2002, Hawkins et al. used 2dF data to test the periodicity in $\log(1+z_{QSO})$ and found no periodicity [18]. However, Napier & Burbidge argued that to test the non-cosmological QSO redshift hypothesis, only late-type active spiral systems should be used as parent galaxies [11]. Arp et al. introduced a new contour method and reexamined the 2dF sample [10]. They claimed that the redshifts of bright QSOs in the QSO density contours fit Karlsson formula and thus confirmed the redshift periodicity. However, as shown in their Figure 3, the peaks are not obvious by eye and there is no statistical results presented to show their significance. We used the SDSS DR1 data to construct the contours defined by Arp et al. [10] and found no evidence for redshift peaks at the predicted positions, as shown in Figure 3 [19]. We also checked the periodicities in QSOs which are paired with nearby active galaxies (therefore possibly ejected by the paired galaxy according to the non-cosmological QSO redshift hypothesis) in several different ways, and found no periodicity [19]. One example is shown in Figure 4.

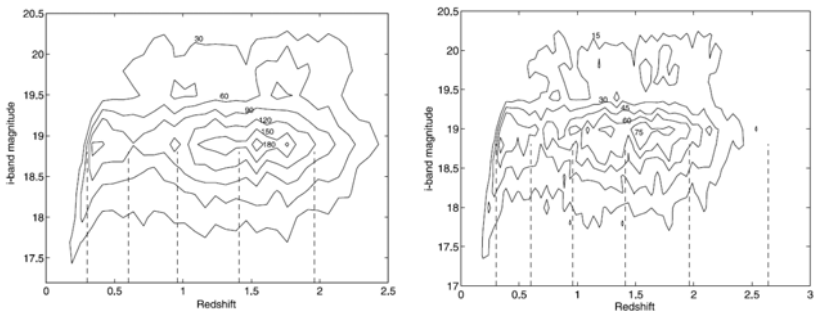

**Figure 3.** Apparent magnitude vs. measured redshift for QSOs in the SDSS DR1 catalog. In the left panel, the whole region is divided into boxes $\Delta z * \Delta B = 0.075 * 0.3$ in the redshift /apparent magnitude plane, while $\Delta z * \Delta B = 0.05 * 0.2$ in the right panel. The contours represent QSO density in steps of 180, 150, 120, 90, 60, and 30 in the left panel, while in the right panel they represent 75, 60, 45, 30, and 15, from the innermost (high density) to outermost (low density). The predicted Karlsson peaks at z = 0.30, 0.60, 0.96, 1.41, 1.96, and 2.64 are shown by vertical lines. (This figure is taken from [19].)

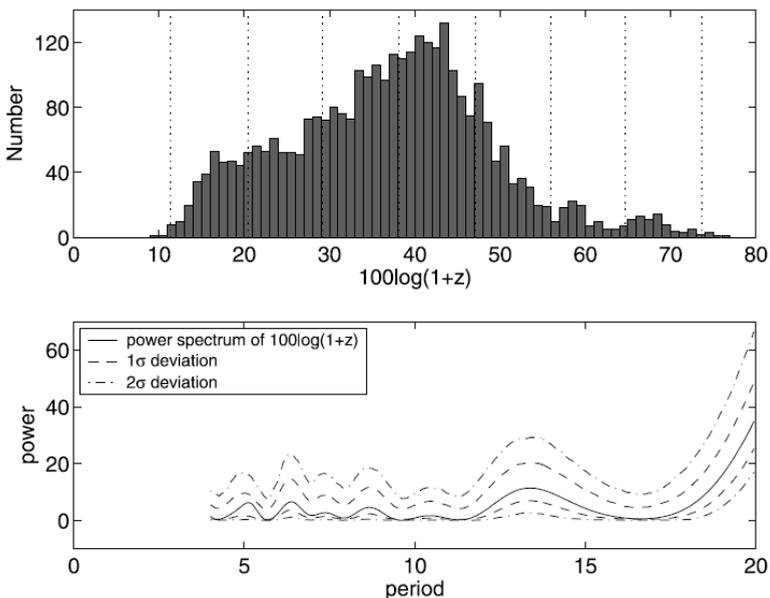

**Figure 4.** Effective redshifts of 3216 QSOs paired with nearby active galaxies (star-forming or starburst galaxies) with a projection distance less than 200 kpc. Top: Histogram of redshifts of these QSOs with peaks predicted by Karlsson's formula indicated by dotted vertical lines. Bottom: Unwindowed power spectra of 100 log (1+z) (solid line) with 1-sigma (dashed lines) and 2-sigma (dash-dotted lines) deviations given from 1000 bootstrap simulations. (This figure is taken from [19].)

The main remaining argument reported as supporting evidence of non-cosmological redshift hypothesis comes from the two dips in the SDSS QSO redshift distribution at z=2.7 and 3.5 [31], which are caused by the reduced efficiency of the selection algorithm at these redshifts [30]. These two dips lead to a strong periodicity around 0.67 in QSO redshifts, as shown in Figure 5 [19]. If only the high-completeness sample is used, the periodicity disappear, as shown in Figure 6(a) [19]. After correcting the observation efficiency, there is no structure in the redshift distribution of SDSS DR5 QSOs, in contrary to the claimed structure in [31], as shown in Figure 7 [20]. Also, there is no such structure in 2dF data, as shown in Figure 8 [19]. In sum, same as discussed in [10], though the periodicity (or dips) in the data is apparent, however, the peaks and dips themselves, for which periodicity is claimed, are caused by selection effects and are not real. Therefore, there is no real periodicity.

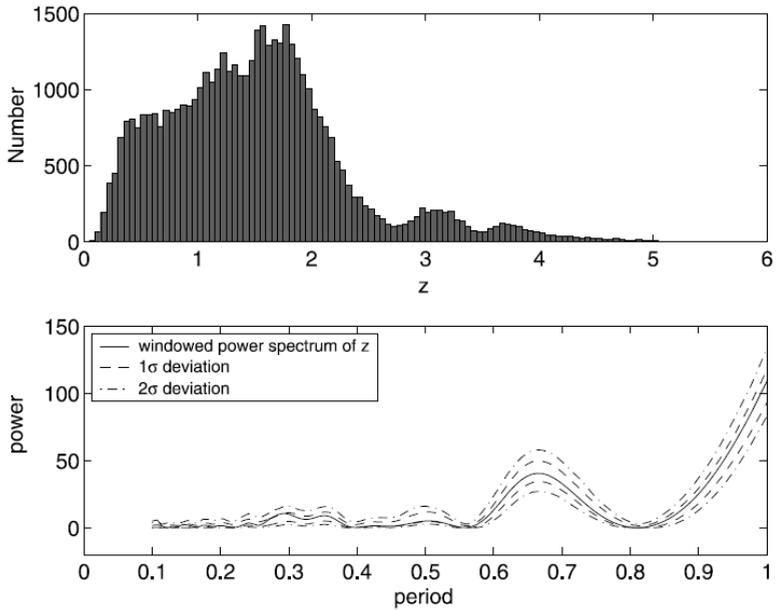

**Figure 5.** Redshifts of 46,420 QSOs in SDSS DR3. Top: Histogram of redshifts of these QSOs. Bottom: Power spectra of z (solid line) weighted using a Hann function with 1-sigma (dashed lines) and 2-sigma (dash-dotted lines) deviations given from 1000 bootstrap simulations. (This figure is taken from [19].)

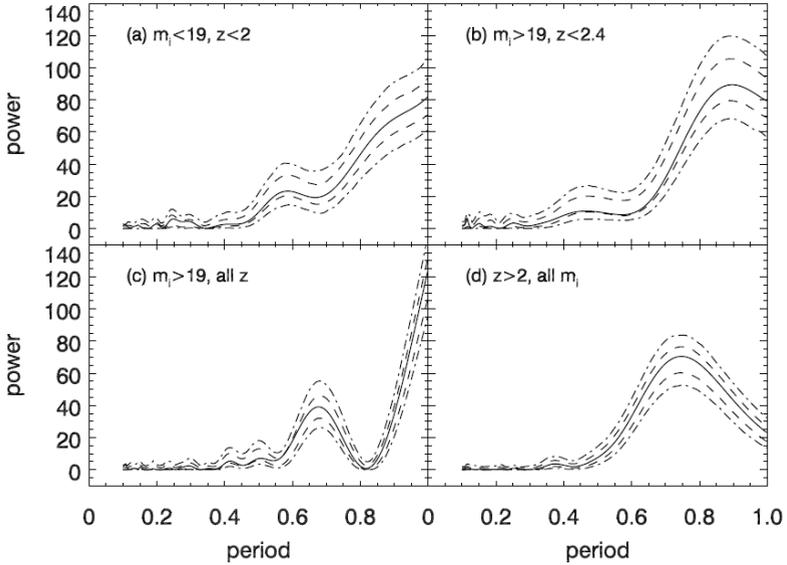

**Figure 6.** Power spectra of redshift of four subsamples from SDSS DR3. Panel (a) is for the high-completeness sample containing 23,109 QSOs with $m_i <19$ and $z < 2$, and the other panels are for samples containing QSOs in low-completeness regions: (b) 15,696 QSOs with $m_i >19$ and $z < 2:4$, (c) 19,064 QSOs with $m_i >19$, and (d) 9763 QSOs with $z > 2$. The power spectra of z (solid line) is weighted using a Hann function with 1-sigma (dashed lines) and 2-sigma (dash-dotted lines) deviations given from 1000 bootstrap simulations. (This figure is taken from [19].)

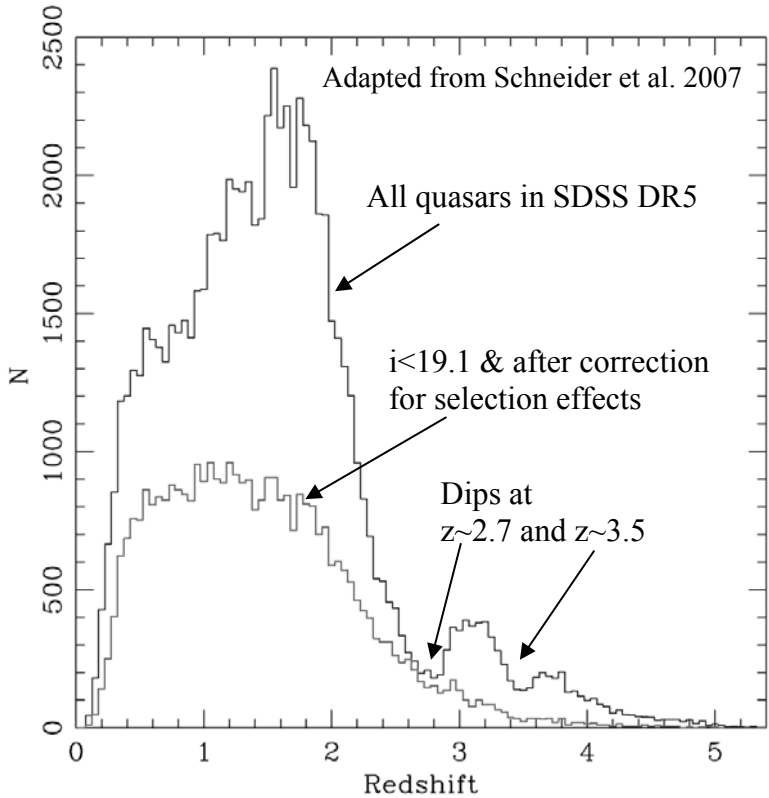

**Figure 7.** Redshift histogram of the cataloged quasars in SDSS DR5. The dips at redshifts of 2.7 and 3.5 are caused by the reduced efficiency of the selection algorithm at these redshifts. The lower histogram is the redshift distribution of the i <19.1 sample after correction for selection effects [20]. (This figure is adapted from [20].)

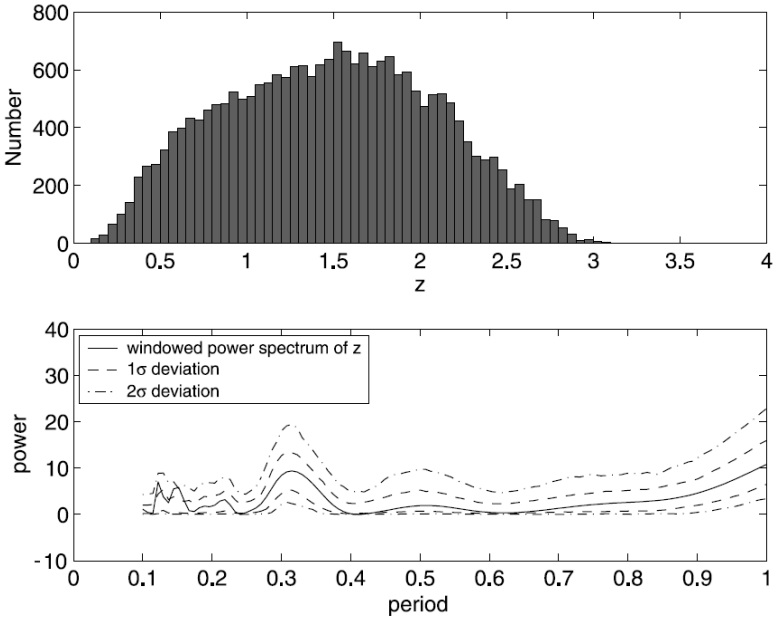

**Figure 8.** Same as Fig. 5, but for redshifts of 22,497 QSOs with the highest quality flag in 2dF. (This figure is taken from [19].)

## 3. Conclusion and Discussion

From the above discussion, we safely reach the conclusion that there is no solid evidence in SDSS and 2dF data supporting the non-cosmological QSO redshift hypothesis. In contrary, there are many solid examples which could only be explained if QSOs are cosmological distant objects. One example is lensed QSOs by foreground galaxies where QSOs must be more distant than the galaxies. Another example is Lyman-alpha forest and Gunn-Peterson trough seen in high-z QSOs, where QSOs must pass through high-z ISM with large integrated neutral hydrogen column densities. Also, if QSOs are ejected by galaxies, we should have seen some QSOs in the field of nearby galaxies with proper motions but we did not.

The mysterious mechanism to produce intrinsic non-cosmological redshifts also makes it less appealing. If the redshift is Doppler shift of ejection velocity, we expect to see a large fraction of blueshifted QSOs but we did not (but see [32] and Chapter 8 of this book). As Occam's razor says, the explanation of any phenomenon should make as few assumptions as possible. In this sense, non-cosmological QSO redshift models are too complex, in comparison with the commonly accepted cosmological QSO redshift model which offers a physically well motivated and yet very simple explanation to the nature of QSOs, and is coherently supported by overwhelming observations in astronomy. We thus conclude that there is neither need, nor soild basis for non-cosmological QSO redshift models.

## Acknowledgements

We thank the editor Dr. Basu for inviting us to contribute to this book. We are grateful to Dr. Schneider for allowing us to include Figure 3 of [20] as Figure 7 in this chapter. SNZ acknowledges partial funding support by the Yangtze Endowment from the Ministry of Education at Tsinghua University, The National 973 Project of the Ministry of Science and technology, Directional Research Project of the Chinese Academy of Sciences under project No. KJCX2-YW-T03 and by the National Natural Science Foundation of China under grant Nos. 10521001, 10733010 and 10725313.